# IRS-Assisted Millimeter-wave Massive MIMO with Transmit Antenna Selection for IoT Networks

Taissir Y. Elganimi[1][0000-0002-2401-6264], Khaled M. Rabie[2][0000-0002-9784-3703] and Galymzhan Nauryzbayev[3][0000-0003-4470-3851]

[1] Department of Electrical and Electronic Engineering, University of Tripoli, Libya
`t.elganimi@uot.edu.ly`
[2] Department of Engineering, Manchester Metropolitan University, United Kingdom
`k.rabie@mmu.ac.uk`
[3] School of Engineering and Digital Sciences, Nazarbayev University, Kazakhstan
`galymzhan.nauryzbayev@nu.edu.kz`

**Abstract.** An intelligent reflecting surface (IRS)-assisted millimeter-wave (mmWave) massive multiple input multiple output (MIMO) system with transmit antenna selection (TAS) using orthogonal space-time block codes (OSTBC) scheme is proposed in this paper. This system combines TAS and IRS with hybrid analog-digital beamforming (HBF) for 60 GHz mmWave communications in order to exploit the benefits of TAS, OSTBC, analog beamforming (ABF), and transmit digital precoding techniques. The proposed system, however, benefits from the transmit diversity gain of OSTBC scheme as well as from the signal-to-noise ratio (SNR) gains of both the beamformer and the IRS technology. The simulation results demonstrate that TAS-OSTBC system with zero-forcing precoding technique outperforms the conventional TAS system with OSTBC scheme. Furthermore, the bit error rate (BER) performance significantly improves as the number of antenna array elements increases due to providing a beamforming gain. In addition, increasing the number of reflecting elements further enhances the error performance. It is also found from the simulation results that the TAS-OSTBC system with hybrid precoding has better BER performance than that of TAS-OSTBC with ABF, and IRS-assisted systems significantly outperform the conventional systems without the IRS technology. This makes the proposed IRS-assisted system an appealing solution for internet-of-things (IoT) networks.

## 1    Introduction

Multiple-input multiple-output (MIMO) systems have improved the capacity and the reliability of wireless communications [1], and enhanced the spectral efficiency of wireless links through physical layer techniques [2], especially, when the number of receive antennas is greater than or equal to the number of transmit antennas [3]. However, due to the practical constraints, employing multiple transmit antennas at the transmitter only is required to achieve good performance by means of space time coding schemes. One of these schemes is the space-time block codes (STBC) which was first proposed by Alamouti in [4] to take advantage of the benefits and potential of MIMO systems due to its low decoding complexity and implementation simplicity. In addition, it was proposed to achieve a full diversity gain with two transmit antennas and only one receive antenna with the same diversity order of maximal ratio combining (MRC) with only one transmit antenna and two receive antennas [3].

The data rate enhancement of MIMO systems is normally associated with extra complexity, size and high hardware cost. This is mainly because a power-hungry radio frequency (RF) chain is required in such systems to be directly connected to the port of each transmit antenna. To overcome this issue, a promising technique known as antenna selection (AS) is considered as a solution to reduce the number of RF chains, where a subset of the transmit and/or receive antennas can be selected and activated for transmission [5]. This leads to effectively reducing the cost and attaining many of the advantages of MIMO systems. One of the well-known AS categories is transmit antenna selection (TAS), where one or multiple antennas are selected to convey information, and achieve full-diversity benefits [6-8]. TAS using orthogonal space-time block codes (OSTBC) is the most well-known TAS scheme that was first proposed by Gore and Paulraj in [6], and experienced rapid progress and increased research focus. However, TAS and OSTBC systems have been combined in order to boost both multiplexing gain and diversity gain simultaneously, and to exploit their advantages. In TAS-OSTBC systems, a practical method for attaining a full-diversity order is considered by transmitting the OSTBC signal matrices over the selected antennas [4, 9]. Therefore, only two transmit antennas are selected for transmission in TAS-OSTBC systems.

In the past few years, the intelligent reflecting surface (IRS) technology has gained immense popularity in improving the energy efficiency of wireless systems by intelligently reconfiguring and controlling the wireless propagation environment. The IRS technology was first proposed in [10] as a promising technology for the future 6G wireless communications and beyond systems in order to address the main two issues in millimeter-wave (mmWave) massive MIMO systems, namely, the high energy consumption and the blockage issue [11]. The IRS technology has great potential to cost-effectively enhance the system performance, to realize a software-controllable and programmable wireless environment via software-controlled reflection [12], and to provide an energy-efficient alternative to the traditional power-hungry phased array [13].

Generally, massive MIMO communications for internet-of-things (IoT) is a developing topic. This is mainly because the IoT connectivity has some requirements that are different from the broadband communications, such as the coverage, bandwidth, power consumption, latency, cost, and data throughput. Therefore, utilizing massive MIMO connections for IoT connectivity is important, and massive IoT is preferred when broadband coverage is needed [14], especially, in reduced RF chain communication systems.

In this paper, a zero-forcing (ZF) precoding technique is employed in TAS with Alamouti STBC scheme, and the bit error rate (BER) performance is evaluated. In this scheme, the analog beamforming (ABF) is also employed and combined with transmit digital precoding into a hybrid beamforming (HBF) regime for the emerging 60 GHz mmWave communications for the sake of further improving the BER performance. In addition, IRS-assisted mmWave massive MIMO with TAS-OSTBC scheme is proposed where transmission occurs through the reflection of IRS using reconfigurable metasurface materials. For the purpose of highlighting the achieved gains, the error performance of the proposed system is compared to that of TAS-OSTBC with HBF without the IRS, and all systems are compared to the conventional TAS system with OSTBC scheme for the sake of fair comparison. Simulation results showed that significant performance improvements are achieved, and signal-to-noise ratio (SNR) gains are obtained by employing ABF and IRS technology. These gains increase as the number of antenna array elements of the beamformers and the number of passive reflecting elements of IRS increase. Due to the achieved gains and benefits of applying ABF, TAS, STBC and IRS, the proposed systems can be significantly exploited in different scenarios of IoT applications. For example, smart homes and factories, where providing high speed for many IoT devices is needed [15].

The rest of the present paper is organized as follows. In Section 2, the system model of the TAS-OSTBC scheme for IRS-assisted mmWave massive MIMO with hybrid analog-digital beamforming is presented. Simulation results and discussions are demonstrated in Section 3, and finally the paper is concluded in Section 4 with some possible research directions for future work.

## 2 Transmit Antenna Selection Using OSTBC Scheme for IRS-Assisted Millimeter-wave Massive MIMO

In this paper, the Frobenius norm-based TAS using OSTBC scheme has been combined with HBF for IRS-assisted mmWave massive MIMO communications in order to attain improved SNR gains achieved by applying ABF and IRS technology in addition to the transmit diversity gain of OSTBC schemes and reducing the number of the power-hungry RF chains at the transmitter. However, the way of designing HBF in this proposed system model is based on combining and applying both ABF and linear transmit digital precoding techniques in the norm-based TAS-OSTBC system. The base station of this system can communicate with the receiver via the reflection of the IRS units that are intelligently controlled by a smart software controller that attached with IRS in order to coordinate the reflecting modes of the IRS units. The system model of the proposed norm-based TAS-OSTBC for IRS-assisted mmWave massive MIMO with HBF is depicted in Fig. 1.

The following subsections describe the system model and the design of the proposed schemes considered in the simulation results of this paper.

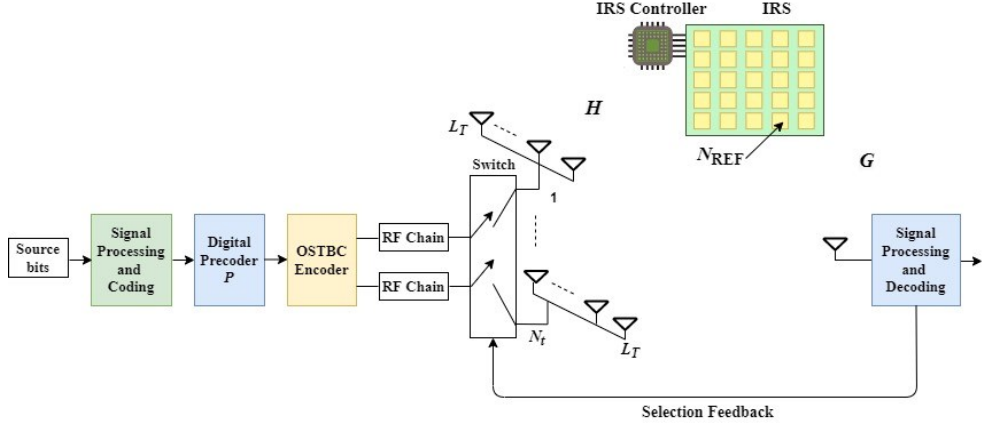

**Fig. 4.** Block diagram of IRS-assisted mmWave massive MIMO system with hybrid analog-digital beamforming and transmit antenna selection using OSTBC scheme.

## 2.1 Norm-Based TAS Using OSTBC Scheme

In multiple antenna systems, transmitting signals from a selected subset of the available transmit antennas is one of the most effective ways that reduces the number of RF chains for the sake of reducing the transmitter cost [3]. Moreover, in order to design a communication system employing STBC schemes under the restrictions of low-cost transmitters, a reliable feedback link connecting the receiver to the transmitter and a clever selection scheme are required where the channel is being perfectly known to the receiver. Therefore, AS is a useful technique where it can be optimized by either maximizing the transmission rate or minimizing the error probability [16]. The latter approach was adopted in this paper, and TAS-OSTBC scheme is considered where only two antennas out of four transmit antennas are selected for conveying OSTBC signal matrices, while the other two transmit antennas are kept silent.

In TAS system with $N_a$ antennas are used among $N_t$ transmit antennas, the effective channel can be represented by $N_a$ columns of $H \in \mathbb{C}^{N_r \times N_t}$. Let $p_i$ denote the index of the $i$-th selected column, where $i = 1, 2, ..., N_a$, then, the corresponding effective channel will be modeled by $N_r \times N_a$ matrix, which is denoted by $H_{p_1, p_2, ..., p_{N_a}} \in \mathbb{C}^{N_r \times N_a}$. The received signal $Y$ of TAS systems can be expressed as [17]

$$Y = \sqrt{\frac{E_s}{N_a}} H_{p_1, p_2, ..., p_{N_a}} X + V, \qquad (1)$$

where $X$ is the space-time coded stream that is mapped into $N_a$ selected transmit antennas, $V$ is the additive noise with identical and independent distributed (i.i.d.) entries having zero mean and variance of $N_o/2$ per dimension (this is the two-sided noise power spectral density), i.e., $\mathcal{CN}(0, N_o/2)$ distribution. The factor $1/\sqrt{N_a}$ ensures that the $N_a$ active antennas share the available symbol energy $E_s$ and that the amount of the total transmitted power used for each channel is independent of the number of active antennas [17, 18]. The $N_a$ antennas can be selected to minimize the upper bound for the pairwise error probability of OSTBC scheme as follows [17]

$$\{p_1^{opt}, p_2^{opt}, ..., p_{N_a}^{opt}\} = \underset{p_1, p_2, ..., p_{N_a} \in A_{N_a}}{\arg \max} \left\| H_{p_1, p_2, ..., p_{N_a}} \right\|_F^2, \qquad (2)$$

where $\|\cdot\|_F$ stands for the Frobenius norm, and $A_{N_a}$ denotes the set of all possible antenna combinations with $N_a$ selected antennas which is defined as

$$A_{N_a} = \binom{N_t}{N_a} = \frac{N_t!}{N_a!\,(N_t - N_a)!}. \tag{3}$$

From (2), the $N_a$ transmit antennas with high column norms are selected to minimize the error rate.

In the case of the TAS-OSTBC scheme with Alamouti STBC encoder, there are only two antennas that are selected and activated for transmission, i.e., $N_a = 2$, and hence the received signal matrix of the norm-based TAS-STBC scheme with two active antennas is written as

$$Y = \sqrt{\frac{E_s}{2}}\, H_{p_1,p_2} X + V. \tag{4}$$

This can be represented as follows

$$[y_1\ \ y_2] = \sqrt{\frac{E_s}{2}}\, [h_1\ \ h_2] \begin{bmatrix} x_1 & -x_2^* \\ x_2 & x_1^* \end{bmatrix} + [v_1\ \ v_2], \tag{5}$$

where $(\cdot)^*$ represents the complex conjugate operation. The linear processing on $Y = [y_1\ \ y_2]$ produces the desired decoupling decision variables for $x_1$ and $x_2$ as follows

$$\tilde{x}_1 = h_1^* y_1 + h_2 y_2^*, \tag{6}$$
$$\tilde{x}_2 = h_2^* y_1 - h_1 y_2^*. \tag{7}$$

## 2.2 TAS-OSTBC Scheme with Zero-Forcing Precoding

Linear transmit precoding in TAS-OSTBC scheme can be implemented with assuming that the channel state information (CSI) is well-known to the transmitter. The total received signal power is maximized with the selection criterion, and hence the received signal matrix of the precoded TAS-OSTBC scheme can be written as follows

$$Y = \sqrt{\frac{E_s}{2}}\, H_{p_1,p_2} \mathcal{P}_D X + V, \tag{8}$$

where $Y \in \mathbb{C}^{N_r \times 2}$ is the complex received signal matrix, $X \in \mathbb{C}^{2 \times 2}$ is the complex transmitted signal matrix, $V \in \mathbb{C}^{N_r \times 2}$ is the additive noise matrix, and $\mathcal{P}_D \in \mathbb{C}^{2 \times 2}$ is the diagonal form of the precoder matrix $\mathcal{P}$ that can be obtained by the knowledge of the channel matrix $H_{p_1,p_2}$ of TAS-OSTBC scheme.

In MIMO systems, ZF linear precoders are typically employed extensively. The main objective of adopting ZF precoders is to eliminate the co-channel interference as well as the inter-symbol interference among the transmit antennas [19]. However, when the number of transmit and receive antennas is equal, ZF precoding can be used to acquire the channel inverse at the transmitter side, but a higher number of transmit antennas than the receive antennas can be used to avoid the problem of raising the total transmit power [20]. In this paper, the transmitter pre-multiplies the STBC transmission matrix $X$ by the diagonal form of the precoder matrix $\mathcal{P}$ which is expressed for TAS-OSTBC systems with ZF precoding scheme as

$$\mathcal{P} = \beta \mathcal{P}_{ZF} = \beta H_{p_1,p_2}^H \left[ H_{p_1,p_2} H_{p_1,p_2}^H \right]^{-1}, \tag{9}$$

where $(\cdot)^H$ and $(\cdot)^{-1}$ represent the Hermitian (conjugate transpose) operation and the inversion operation, respectively, $\mathcal{P}_{ZF}$ is the ZF precoding matrix without power scaling, and $\beta$ is the scaling factor which is a constant used to meet the total transmit power constraint after implementing the transmit digital precoding, and it is given as follows [17, 21]

$$\beta = \sqrt{\frac{N_t}{trace\ (\mathcal{P}_{ZF}\mathcal{P}_{ZF}^H)}}. \tag{10}$$

In the precoded TAS-OSTBC systems, the received signal is normalized in order to meet the total transmit power constraint after implementing this linear precoder by dividing the received signal matrix by $\beta$ through a receiver's automatic gain control (AGC) [17].

Since there are only two active antennas in the norm-based TAS-OSTBC scheme, the effective channel matrix $H_{p_1,p_2}$ has the size of $N_r \times 2$, and hence the precoder matrix $\mathcal{P}$ becomes with size of $2 \times N_r$. In this paper, TAS-OSTBC systems are considered with one receive antenna, thus the complex-valued precoder $\mathcal{P} \in \mathbb{C}^{2\times 1} = \begin{bmatrix} p_1 \\ p_2 \end{bmatrix}$ can be obtained from Equation (9) where the effective channel can be expressed as $H_{p_1,p_2} \in \mathbb{C}^{1\times 2} = [h_1\ h_2]$. Therefore, the diagonal and complex-valued matrix $\mathcal{P}_D \in \mathbb{C}^{2\times 2}$ is expressed as

$$\mathcal{P}_D = \text{diag}(\mathcal{P}) = \begin{bmatrix} p_1 & 0 \\ 0 & p_2 \end{bmatrix}. \tag{11}$$

Then, the received signal of Equation (8) can be represented as follows

$$\begin{aligned}[] [y_1\ y_2] &= \sqrt{\frac{E_s}{2}}\ [h_1\ h_2] \begin{bmatrix} p_1 & 0 \\ 0 & p_2 \end{bmatrix} \begin{bmatrix} x_1 & -x_2^* \\ x_2 & x_1^* \end{bmatrix} + [v_1\ v_2] \\ &= \sqrt{\frac{E_s}{2}}\ [h_1 p_1\ h_2 p_2] \begin{bmatrix} x_1 & -x_2^* \\ x_2 & x_1^* \end{bmatrix} + [v_1\ v_2]. \end{aligned} \tag{12}$$

The following linear processing on $Y = [y_1\ y_2]$ produces the desired decoupling decision variables for both $x_1$ and $x_2$ with taking the real-valued effective channel $H_P \in \mathbb{C}^{1\times 2} = H_{p_1,p_2}\mathcal{P}_D = [h_1\ h_2]\begin{bmatrix} p_1 & 0 \\ 0 & p_2 \end{bmatrix} = [h_1 p_1\ h_2 p_2]$ into consideration as follows

$$\tilde{x}_1 = h_1^* p_1^* y_1 + h_2 p_2 y_2^*, \tag{13}$$
$$\tilde{x}_2 = h_2^* p_2^* y_1 - h_1 p_1 y_2^*. \tag{14}$$

## 2.3 TAS-OSTBC Scheme with Analog Beamforming

In this paper, implementing ABF in TAS-OSTBC systems is presented in order to improve the BER performance as the number of antenna elements increases. In this scheme, analog pre-processing network is used at the transmitter of TAS-OSTBC systems. This network cancels the interference and improves the system with reducing the number of RF chain components such as analog-to-digital converters (ADCs), mixers, power amplifiers, synthesizers and filters [22]. In addition, the mmWave massive MIMO channel is considered, and it is assumed that ABF arrays are used at the transmitter side, and the receiver has a perfect CSI knowledge in a quasi-static flat Rayleigh fading channel.

The ABF is controlled in this paper based only on the angle of departure (AoD) $\theta_{AoD}$ at the transmitter. More specifically, when adopting uniform linear array (ULA) antennas, the ABF weight $w_{T_i}$ with $L_T$ elements at each transmit antenna is modeled as follows [23, 24]

$$w_{T_i} = \frac{1}{\sqrt{L_T}} \begin{bmatrix} 1 & e^{j\delta_T^{(i)}} & e^{j2\delta_T^{(i)}} & \ldots\ldots\ldots & e^{j(L_T-1)\delta_T^{(i)}} \end{bmatrix}^T, \tag{15}$$

where $(\cdot)^T$ represents the transpose operation, $\delta_T^{(i)}$ is the electrical phase shift between each two adjacent antenna array elements along the transmit antenna that is presented as $kd\sin(\theta_{AoD}^{(i)})$, $\theta_{AoD}^{(i)}$ denotes the AoD towards the $i$-th ABF of the receiver, $k$ is the wavenumber ($=2\pi/\lambda$), where $\lambda$ is the corresponding transmission wavelength, and $d$ is the spacing between each pair of antenna elements in each ABF with the constraint of $d \leq \lambda/2$ in order to achieve a beneficial beamforming gain [25].

These ABF weights are known as the ABF weight vectors with $L_T$ complex conjugate coefficients at the transmitter, and these vectors contain the information about all antenna elements and the direction of arrivals (DoA) of the transmitted signals.

By considering that each transmit antenna in TAS-OSTBC system has an array antenna of $L_T$ antenna elements that are uniformly spaced, this scheme is based on using the weight vector in Equation (15). Thus, the received signal matrix of TAS-OSTBC scheme with ABF can be expressed as

$$Y = \sqrt{L_T} \sum_{i=0}^{L_T-1} \sqrt{\frac{E_s}{2}} w_{T_i}^H H_{p_1,p_2} w_{T_i} X + V. \tag{16}$$

One key issue in most of the beamforming algorithms is the normalization technique. In implementing ABF in the norm-based TAS-OSTBC aided mmWave massive MIMO system, the received signal is normalized by the factor $\sqrt{L_T}$ which ensures that the received signal power scales linearly with the number of antenna elements $L_T$. Therefore, the received signal equation is divided by the factor $\sqrt{L_T}$. In this case, the output power of TAS-OSTBC-ABF system with $L_T$ transmit antenna elements becomes equal to the output power in the case of single element antennas [26]. In addition, the effective channel matrix $\sqrt{L_T} \sum_{i=0}^{L_T-1} w_{T_i}^H H_{p_1,p_2} w_{T_i}$ is taken into consideration for the linear processing at the receiver side of TAS-OSTBC aided mmWave massive MIMO system.

## 2.4 Hybrid Beamforming Design in TAS-OSTBC Scheme

Combining HBF with Frobenius norm-based TAS system using OSTBC scheme can be done by applying transmit ABF in the precoded TAS-OSTBC system with assuming that the knowledge of the channel matrix is available at the transmitter side of TAS-OSTBC-ABF system which is previously presented. In this paper, ZF precoding technique is used with transmit ABF to design the TAS-OSTBC-HBF scheme for a mmWave massive MIMO system. The received signal of TAS-OSTBC-HBF scheme is expressed as

$$Y = \sqrt{L_T} \sum_{i=0}^{L_T-1} \sqrt{\frac{E_s}{2}} w_{T_i}^H H_{p_1,p_2} w_{T_i} \mathcal{P}_D X + V. \tag{17}$$

In TAS-OSTBC scheme aided mmWave massive MIMO system with HBF, both number of transmit antenna elements and the constant $\beta$ are taken into account to normalize the received signal with a factor of $\beta\sqrt{L_T}$. In addition, the effective channel matrix $\sqrt{L_T} \sum_{i=0}^{L_T-1} w_{T_i}^H H_{p_1,p_2} w_{T_i} \mathcal{P}_D$ is taken into consideration for the linear processing at the receiver side.

## 2.5 TAS-OSTBC Scheme for IRS-Assisted mmWave Massive MIMO System

In the proposed system model of TAS-OSTBC scheme for IRS-aided mmWave massive MIMO, the properties of the low-cost IRS reflecting elements can be expressed by the diagonal phase-shift matrix as follows [27, 28]

$$\Phi = \alpha \operatorname{diag}(e^{j\theta_1}, e^{j\theta_2}, \ldots\ldots e^{j\theta_{N_{\text{REF}}}}), \tag{18}$$

where $\alpha \in (0,1]$ is the fixed amplitude coefficient, $N_{\text{REF}}$ is the number of reflecting elements, and $\theta_1, \theta_2, \ldots\ldots, \theta_{N_{\text{REF}}}$ are the IRS phase-shifts for each passive element. This diagonal reflection matrix accounts for the effective IRS phase shifts applied by all IRS elements.

The received signal of the proposed IRS-assisted mmWave massive MIMO system with TAS-OSTBC scheme and hybrid precoding can be expressed as

$$Y = \sqrt{L_T} \sum_{i=0}^{L_T-1} \sqrt{\frac{E_s}{2}} w_{T_i}^H H_{eff_{p_1,p_2}} w_{T_i} \mathcal{P}_D X + V, \tag{19}$$

where $H_{eff} = \sum_{r=1}^{N_{\text{REF}}} G\Phi H$ is the effective channel matrix with the presence of single IRS, $r = 1, 2, \ldots, N_{\text{REF}}$, $H_{eff_{p_1,p_2}} \in \mathbb{C}^{N_r \times 2}$ is the effective channel represented by $N_a = 2$ columns of $H_{eff} \in \mathbb{C}^{N_r \times N_t}$, $G \in \mathbb{C}^{N_r \times N_{\text{REF}}}$ is the channel matrix between the single IRS and the $N_r$ receive antennas, and $H \in \mathbb{C}^{N_{\text{REF}} \times N_t}$ is the channel matrix between the IRS and the transmitter of the proposed IRS-assisted system. In this paper, $H$ and $G$ are modeled as Rayleigh fading channels with $\mathcal{CN}(0,1)$ distribution, i.e., with zero mean and unit variance per dimension. Similarly, the effective channel matrix $\sqrt{L_T} \sum_{i=0}^{L_T-1} w_{T_i}^H H_{eff_{p_1,p_2}} w_{T_i} \mathcal{P}_D$ is taken into consideration for the linear processing at the receiver side of the proposed IRS-assisted mmWave massive MIMO system with TAS-OSTBC scheme.

## 3   Simulation Results and Discussions

In order to evaluate the BER performance of TAS-OSTBC-HBF for IRS-assisted mmWave massive MIMO systems, intensive simulations have been performed in MATLAB environment. The results provided by TAS-OSTBC-ABF systems are compared to the state-of-the-art Alamouti's STBC transmission scheme for a $2 \times 1$ system and single input single output (SISO) scheme. Throughout the simulation, the performance results of the proposed systems employing 4-ary quadrature amplitude modulation (QAM) technique with 4 transmit antenna arrays and 1 receive antenna are presented, while using the simulation parameters outlined in Table 1. In addition, $\alpha = 1$ is assumed in the proposed IRS-assisted systems.

**Table 1.** Simulation parameters.

| Parameters | Values |
| --- | --- |
| Modulation scheme | 4-QAM |
| Number of frames per packet | 100000 |
| Number of packets | 10 |
| Number of transmit antennas | 4 |
| Number of receive antennas | 1 |
| Number of array elements | $L_T$ |
| Channel models | Rayleigh Fading |
| Transmission wavelength | $\lambda = 0.5$ cm |
| Carrier frequency | 60 GHz |
| Antenna spacing | $\lambda/2$ |

The BER performance versus SNR of TAS-OSTBC system with ABF is shown in Fig. 2 with different numbers of antenna array elements using ULA antennas. This figure shows that the conventional TAS-OSTBC system outperforms SISO and Alamouti STBC schemes, and a diversity gain has been achieved by using the same number of RF chains in both TAS-OSTBC and Alamouti systems. It is also demonstrated that an error performance improvement is achieved by increasing the number of beam-steering elements of each transmit antenna array of TAS-OSTBC system. Furthermore, a performance improvement is achieved and SNR gains of approximately 3 dB, 6 dB, 9 dB, 12 dB and 15 dB are obtained by TAS-OSTBC-ABF schemes with $L_T = 2, 4, 8, 16$ and $32$ elements, respectively, as compared to the conventional TAS-OSTBC scheme with single element antennas at the transmitter side.

In Fig. 3, the BER performance of TAS-OSTBC with ZF precoding technique is depicted and showed that there is an improvement of about 3.4 dB over the conventional TAS-OSTBC scheme. Moreover, the BER performance of TAS-OSTBC system with HBF showed an SNR improvement of almost 3 dB, 4.7 dB, 6 dB, 9 dB, 12 dB and 15 dB with $L_T = 2, 3, 4, 8, 16$ and $32$ array elements, respectively, over the ZF precoded TAS-OSTBC system with single element antennas. Additionally, the SNR gain of ABF can be approximately expressed as $10 \log_{10} L_T$ in decibel. This amount is the SNR improvement that achieved at a particular BER performance by employing

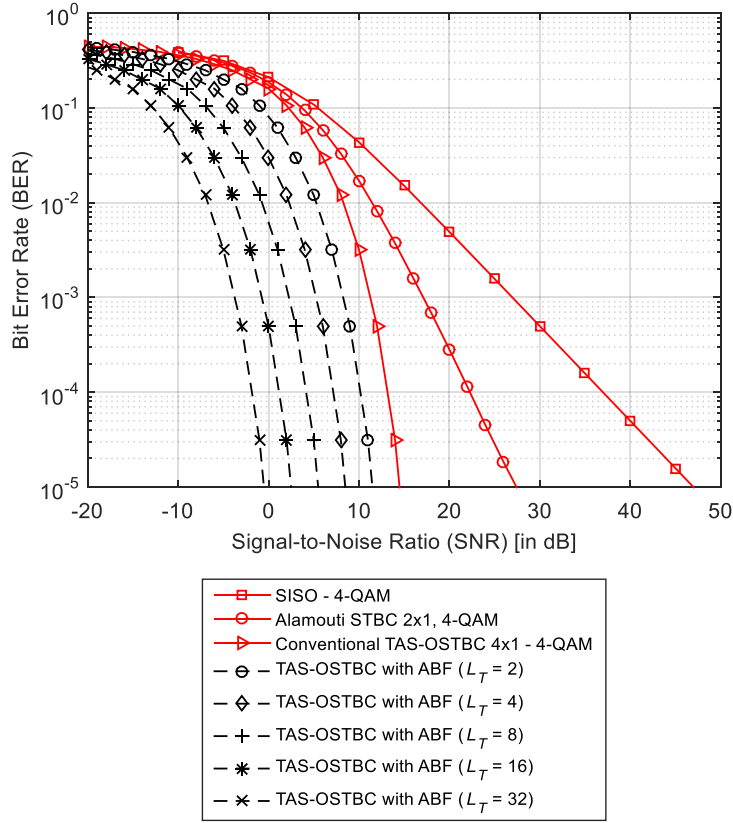

**Fig. 2.** BER performance of TAS-OSTBC scheme employing 4-QAM technique with analog beamforming.

ABF with $L_T$ array elements at each transmit antenna in TAS-OSTBC and ZF precoded TAS-OSTBC systems as compared to the conventional systems with single element antennas at the transmitter.

Fig. 4 shows the BER performance of both TAS-OSTBC-ABF and TAS-OSTBC-HBF schemes versus the number of transmit antenna array elements for different SNR values. It is obvious from this figure that TAS-OSTBC-HBF systems outperform TAS-OSTBC-ABF systems in all cases. On the other hand, the BER performance of TAS-OSTBC-HBF systems is better than that of TAS-OSTBC-ABF with the same number of array elements, i.e., the number of phase shifters that are required at the transmitter side. It is also noticeable from this figure that the required number of array elements at each transmit antenna in TAS-OSTBC-ABF schemes is just more than twofold the required number of antenna elements in TAS-OSTBC-HBF scheme to achieve the same BER performance.

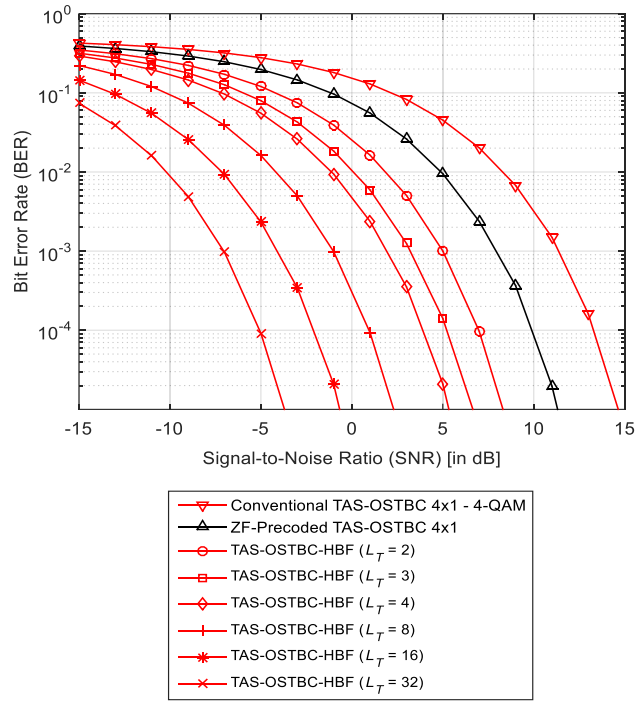

**Fig. 3.** BER performance of TAS-OSTBC scheme with hybrid analog-digital beamforming employing 4-QAM technique.

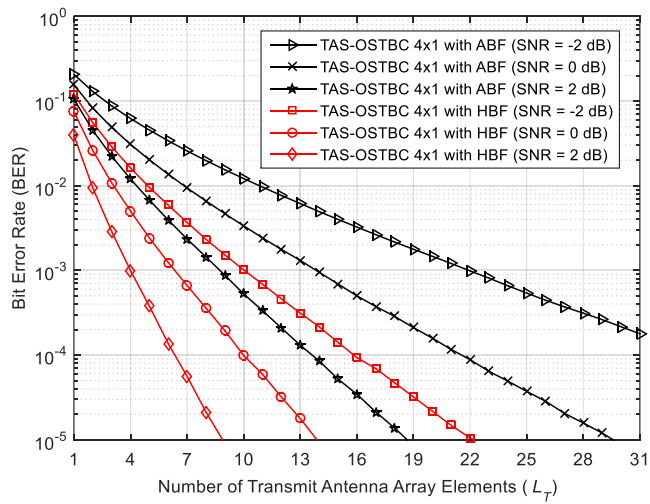

**Fig. 4.** BER performance versus the number of array elements in TAS-OSTBC-ABF and TAS-OSTBC-HBF schemes employing 4-QAM technique.

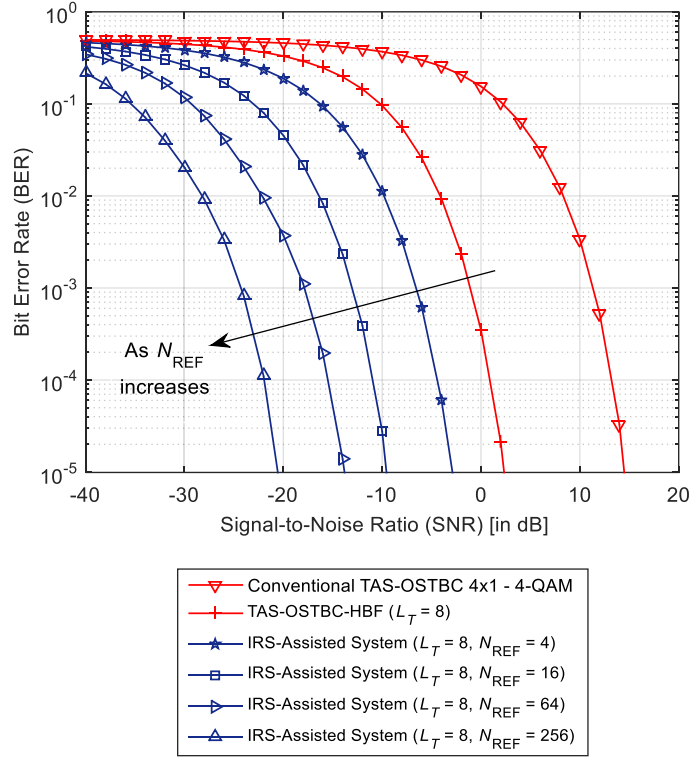

**Fig. 5.** BER performance of IRS-assisted mmWave massive MIMO systems with TAS-OSTBC scheme and hybrid analog-digital beamforming employing 4-QAM technique.

Finally, the BER performance versus SNR of IRS-assisted mmWave massive MIMO system with TAS-OSTBC scheme is shown in Fig. 5 with different numbers of reflecting elements and 8 antenna array elements. It is clear from this figure that the BER performance of the IRS-assisted systems is better than that of TAS-OSTBC-HBF. It is also clear that the BER performance improves as the number of reflecting elements increases. In addition, it is found that there is an SNR gain of about 6 dB, 12 dB, 18 dB and 24 dB with $N_{\text{REF}} = 4$, 16, 64 and 256 passive elements, respectively. This SNR gain that achieved by deploying a single IRS can be approximately expressed as $10 \log_{10} N_{\text{REF}}$ in decibel.

## 4    Conclusions

In this paper, IRS-assisted mmWave massive MIMO with TAS-OSTBC and hybrid analog-digital beamforming is proposed where ZF precoding scheme and ABF arrays at the transmitter are employed in order to design HBF and exploit the transmit diversity potential of MIMO channels. Thereby, it has been revealed that employing ABF provides a beamforming gain, and the achievable BER performance can be considerably improved as the number of antenna array elements increases. Additionally, the obtained results of TAS-OSTBC-HBF systems showed a better error performance than that of TAS-OSTBC-ABF systems with the same number of antenna array elements in each transmit antenna. This performance enhancement of almost 3.4 dB is a result of using the ZF digital precoders in TAS-OSTBC systems, and hence a better BER performance is attained in ZF precoded TAS-OSTBC system as compared to the conventional TAS-OSTBC scheme. In addition, IRS-assisted mmWave massive MIMO systems with TAS-OSTBC showed that significant BER improvements can be achieved, and the performance can be substantially improved as the number of reflecting elements increases. Furthermore, only two power-hungry RF chains are required in these systems, thus the synchronization of all transmit antennas in the proposed systems is not needed. To conclude, IRS-assisted mmWave massive MIMO systems with TAS-OSTBC can be useful for the future 6G communication systems. It can be also regarded as a promising candidate to be exploited for different scenarios in IoT applications. Our future studies will focus on deploying multiple IRSs and multiple users with imperfect CSI scenario.